# Are high school degrees and university diplomas equally heritable in the US? A new measure of relative intergenerational mobility

Anna Naszodi[♦] and Liliana Cuccu[•]

This paper proposes a new measure of relative intergenerational mobility along the educational trait as a proxy of inequality of opportunity. The new measure is more suitable for controlling for the variations in the trait distributions of individuals and their parents than the commonly used intergenerational persistence coefficient. This point is illustrated by our empirical analysis of US census data from the period between 1960 and 2015: we show that controlling for the variations in the trait distributions adequately is vital in assessing the part of intergenerational mobility which is not caused by the educational expansion. Failing to do so can potentially reverse the relative priority of various policies aiming at reducing the "heritability" of high school degrees and tertiary education diplomas.

## 1. Introduction

This paper applies a new measure of relative intergenerational mobility along the educational trait in order to quantify inequality of opportunity. This measure is able to control for variations in the educational distribution of young adults as well as that of their parents. It

---

[♦] Anna Naszodi, corresponding author, email: anna.naszodi@gmail.com. She is an honorary member of the KRTK Centre for Economic and Regional Studies, Institute of Economics.

[•] Liliana Cuccu, Universitat de Barcelona.

was developed by Liu and Lu (2006), who originally used it for assessing the degree of sorting in marriages. They show analytically that certain conventional measures of the degree of sorting cannot control for variations in the education levels of marriageable men and marriageable women, while their measure can do.

What motivates "importing" the Liu–Lu measure to the literature on mobility is that one of the most commonly applied mobility measures suffers from the same flaw addressed by Liu and Lu. This conventional measure is the intergenerational persistence coefficient (IGPC), which is calculated by regressing individuals' qualification level on the education of their parents and a constant (see e.g. Blanden 2013; Hertz et al. 2007).

Similar to the popular measures applied in the literature of assortative mating, the IGPC is sensitive to variations in the trait distributions. This sensitivity should not be viewed as a weakness as long as the IGPC is interpreted as a measure of linear dependence of individuals' educational attainments on their parents' education levels in a given country at a given point in time. However, the IGPC is often used in cross-country and inter-temporal comparisons even when the educational distributions are not the same across countries or when they change over time. Moreover, this conventional measure of intergenerational mobility often forms the basis of policy recommendations, whereby it is used as a proxy for inequality of opportunity due to the proximity of these concepts (see Narayan et al. 2018).

We think that the Liu–Lu measure adapted for mobility is more suitable to be used for cross-sectional and inter-temporal comparisons of inequality of opportunities associated to parents' education for the following theoretical and empirical reasons.

First, as it is shown by Liu and Lu (2006), their measure adequately accommodates cross-sectional and time series variations in the trait distributions of wives and husbands. Therefore, the "imported" Liu–Lu measure capturing intergenerational mobility does the same with variations in the trait distributions of young adults and their parents.

Second, as we will see, our cross-validation in Section 4 also supports the application of the Liu–Lu measure. For that analysis, we use data on test scores of American students with different family socio-economic status (SES). These data are used to calculate neither the IGPC, nor the Liu–Lu measure. We find the trend in the Liu–Lu measure to be consistent with the closure of the achievement gaps between American low-SES students and high-SES students in some generations born after the Second World War. By contrast, the corresponding dynamics of the IGPC is not consistent with the dynamics of the achievement gaps.

Our empirical analysis exemplifies not only that the two measures subject to comparison can have different dynamics over time. It also shows that these measures can result in different rankings when applied in cross-sectional analyses. The detected difference in rankings means that the Liu–Lu measure and the IGPC disagree on the relative "heritability" of high school degrees and tertiary level diplomas in the US.

This paper is structured as follows. First, we introduce the Liu–Lu measure adapted for capturing intergenerational mobility. Then we apply it, together with the IGPC, to study the patterns of inequality of opportunities in the US using census data from the period between 1960 and 2015. Finally, we validate the Liu–Lu measure as a metric of inequality of opportunity.

## 2. The adapted Liu–Lu measure

This section briefly introduces the Liu–Lu measure adapted to capture inequality of opportunity. In the Liu and Lu (2006) model, the educational trait is described by a dichotomous variable. It can take two possible values, low (L) and high (H). Accordingly, any young adult-parent pair can be of four types. The matrix reporting the number of young adult-parent pairs of each type in a population is the following 2-by-2 matrix:

$$K = \begin{bmatrix} N_{L,L} & N_{L,H} \\ N_{H,L} & N_{H,H} \end{bmatrix}. \qquad (1)$$

If matrix K is known, the trait distribution of the young adults as well as the trait distribution of their parents are known (where the latter is weighted by the number of children). Specifically, the number of young adults with low education level is $N_{L,\cdot} = N_{L,L} + N_{L,H}$ and with high education level is $N_{H,\cdot} = N_{H,L} + N_{H,H}$. The number of young adults with low educated parents is $N_{\cdot,L} = N_{L,L} + N_{H,L}$ and with high educated parents is $N_{\cdot,H} = N_{L,H} + N_{H,H}$. The educational distributions are assumed to be non-degenerate: there is at least one low educated young adult and at least one high educated young adult in the population and also, there is at least one parent in both educational groups.

The Liu–Lu measure is given by:

$$LL(K) = \begin{cases} \frac{N_{H,H} - Q^-}{\min(N_{H,\cdot},\, N_{\cdot,H}) - Q^-}, & \text{if } N_{H,H} \geq Q, \\ \frac{N_{H,H} - Q^+}{Q^+ - \max(0,\, N_{H,\cdot} - N_{\cdot,L})}, & \text{if } N_{H,H} < Q, \end{cases} \qquad (2)$$

where N denotes the total number of young adults, while $Q = N_{H,\cdot} N_{\cdot,H}/N$ is the expected number of high educated young adults with high educated parents under the random allocation of credentials. Furthermore, $Q^-$ is the biggest integer being smaller than or equal to Q, while $Q^+$ is the smallest integer being larger than or equal to Q.

Since high educated people tend to have high educated parents, the empirically relevant cases are those, where there are more H,H-type young adult-parent pairs then would be under the random allocation scenario. Once $N_{H,H} > Q$ is assumed, the Liu–Lu measure simplifies to:

$$LL^{sim}(K) = \frac{N_{H,H} - Q^-}{\min(N_{H,\cdot},\, N_{\cdot,H}) - Q^-}. \qquad (3)$$

The simplified Liu–Lu measure is the same as the Coleman-index defined by Equation (15) in Coleman (1958). It is read as the "actual minus expected over maximum minus minimum", where $N_{H,H}$ is the actual number of H,H-type pairs, while $\min(N_{H,.}, N_{.,H})$ is its maximum value for given row totals and column totals. Finally, $Q^-$ is both the expected (integer) value of the number of H,H-type pairs under random allocation of credentials, and the minimum value of the number of H,H-type pairs if more educated parents tend to have more educated children in all societies.

This metric is scaled:

$$\mathrm{LL}^{\mathrm{sim}}(K) = \begin{cases} 1, & \text{if} \quad K = K^p, \\ 0, & \text{if} \quad K = K^r, \\ \in [0,1], & \text{otherwise,} \end{cases}$$

where $K^p$, and $K^r$ are two benchmark outcomes.

(i) $K^p$ is the maximally unequal outcome, where young adults with low educated parents could obtain a degree only if all the young adults with highly educated parents have a degree:

$$\mathrm{vec}(K^p) = \left[\min(N_{L,.}, N_{.,L}), \max(0, N_{H,.} - N_{.,H}), \max(0, N_{.,H} - N_{H,.}), \min(N_{H,.}, N_{.,H})\right]^T.$$

(ii) Whereas $K^r$ is the perfectly random allocation of resources and talent necessary to obtain a degree: $\mathrm{vec}(K^r) = \left[\frac{N_{L,.} N_{.,L}}{N}, \frac{N_{H,.} N_{.,L}}{N}, \frac{N_{L,.} N_{.,H}}{N}, \frac{N_{H,.} N_{.,H}}{N}\right]^T$.

## 3. The empirical comparison of the Liu–Lu measure with the IGPC

For the empirical analysis, we use the international version of Integrated Public Use Microdata Series (IPUMS) from the Minnesota Population Center. IPUMS International provides data on a large and representative sample covering 1% of the US population for the years 1960, 1970, 2005, 2010 and 2015; 5% for 1980, 1990 and 2000. Our census wave specific samples cover young people aged 30 to 40 years in the respective census years. We use the variable on the reported highest educational attainments of the respondents (edattain), their mother (edattain$_{mom}$) and their father (edattain$_{pop}$).

We compare empirically the Liu–Lu measure and the IGPC. The evolution over time of the former measure is presented by Figure 1, while the evolution over time of the latter measure is presented by Figure 2. The examined indices support the view that the chances of achieving certain credentials were not independent of the parental educational background in the US over the analyzed 55 years: neither the Liu–Lu measure, nor the IGPC takes values close to zero no matter whether parental education corresponds to mother's education level or father's educational attainment. In addition, this finding is also robust to whether high level of education is defined as having at least a high school degree or at least a college degree.

After documenting that high school degrees and tertiary level diplomas, or the skills and resources to obtain them, were to some extent "inherited" in the US, let us turn to examining whether different credentials are equally "heritable" or not. Figures 1, and 2 show that the Liu–Lu measure and the IGPC provide different answers to this question since they generate different rankings with respect to the inequality of opportunities associated to parents' education. Specifically, if the studied inequality of opportunity is proxied by the Liu–Lu measure, then it is higher when the examined opportunity relates to completing high school compared to the case when it is related to achieving a tertiary education (see the relative position of the dashed and the solid lines of Fig.1).

This finding suggests that policies aiming at equalizing the chances of American students with lower parental education to finish high school targeted a higher inequality of opportunity over the analyzed period than policies addressing the same kind of inequality of opportunities to obtain a college degree. Surprisingly, one would draw exactly the opposite conclusion, if the inequality of opportunity were quantified by the IGPC (see the relative position of the dashed and the solid lines of Fig.2).

## 4. A supplementary analysis

To study whether the IGPC or the Liu–Lu measure provides us with a more realistic view, we use a third measure. It is the achievement gap, i.e., the difference in the average test scores of American students from affluent and poor families.

Our focus is on three cohorts: those who were born in the 1950s, 1960s, and 1970s. The members of the first cohort became all part of the age group of young adults in 1990, where young adults are defined as in the previous section, being between 30 and 40. Whereas those in the second cohort were aged between 30 and 40 in 2000. Finally, those in the third cohort reached this age category in 2010.

The key message of Fig.3 is that the achievement gaps (i.e., the 90–10 achievement gap and the 75–25 achievement gap) between those in the top and bottom of the socioeconomic distribution were decreasing in a specific period: both gaps were smaller for the 1970s birth cohort than for the 1960s birth cohort, and those were smaller for the 1960s birth cohort than for the 1950s birth cohort. For instance, the 90–10 gap is around 1.15 standard deviations for the 1954 birth cohort and close to 1.00 standard deviation for the 1978 birth cohort.

One standard deviation is approximately the difference in the average performance of students in 4th and 8th grades. On the basis of this, we can interpret the decline of 0.15 standard deviation over 24 years as a reduction of the gap by at least half a year' worth of learning to 4 years' worth of learning. Although this closure of the achievement gap was far not rapid, its economic significance is indisputable.

In contrast to the achievement gap, the IGPC has not decreased across the same cohorts (see the dashed lines of Figure 2 between 1990 and 2000 and also between 2000 and 2010). According to the IGPC, inequality of opportunity of completing high school has hardly changed, or slightly increased over the decade preceding the turn of the millennium and over the decade following it.

However, the trend of the Liu–Lu measure is in line with the negative trend of the achievement gap (see the dashed lines of Figure 1 between 1990 and 2000 and also between 2000 and 2010). Therefore, our cross-validation corroborates the application of the Liu–Lu measure for quantifying inequality of opportunity.

5. Conclusion

This paper adapted a measure of relative intergenerational mobility from the assortative mating literature. We claim that this new measure is more suitable to capture "how the cake is sliced up" for young people with different parental education than a commonly used measure, the IGPC. In this paper, we presented a theoretical argument supporting our claim. We argued that the new measure is free from the effects of the increases (or decreases) of the cake, i.e., changes in the educational distribution, while the conventional measure is not.

We presented an empirical argument as well by applying a third measure. This third measure is the difference between the test scores of high-SES students and low-SES students aged 14 to 17 years. This metric is independent of certain factors determining the "size of the cake" by construction. For instance, it is affected neither by the number of places offered by universities and colleges, nor by the cutoff score separating students who pass from those who fail the exit exams in high school. We found the dynamics of this third measure to resemble more to the dynamics of the new measure than to that of the commonly used measure.

We applied the new measure together with the conventional measure to study inequality of opportunities associated to parents' education between 1960 and 2015. We found some evidence showing that the difference between the measures compared is empirically relevant. In particular, depending on the measure used, one can form a completely different view as to the question descendants of which educational group were exposed to a higher inequality of opportunities. According to the new measure promoted in this paper, these individuals are from families where the parents have no high school degree while the examined opportunity relates to completing high school. By contrast, the conventional measure reports that the children of parents without a college degree have been suffering from a higher inequality of opportunity for obtaining a tertiary level diploma. This sensitivity highlights the importance of carefully selecting the metrics used for policy making.

# Acknowledgments

The authors acknowledge comments from Lili Márk, as well as from the participants of the internal seminars at the Joint Research Centre of the European Commission. The authors are grateful to Eric A. Hanushek and his coauthors for allowing them to replicate and use one of the figures in their research paper.

# Disclosure statement

# Supplemental material

The code and the data used for the empirical analysis in this paper are published on Mendeley (see: DOI: 10.17632/9bm6wvtfsh ). An earlier version of the paper was published as a preprint (see Naszodi and Cuccu, 2019). The appendix presents the results of the empirical analysis for Austria, France, Greece, Hungary, Portugal, Romania, and Spain over the period between 1968 and 2011.

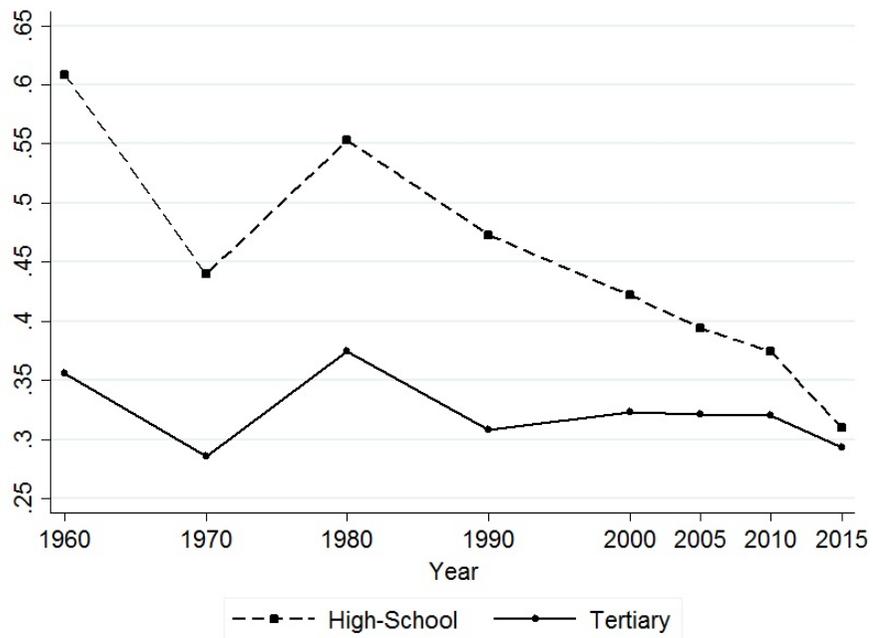 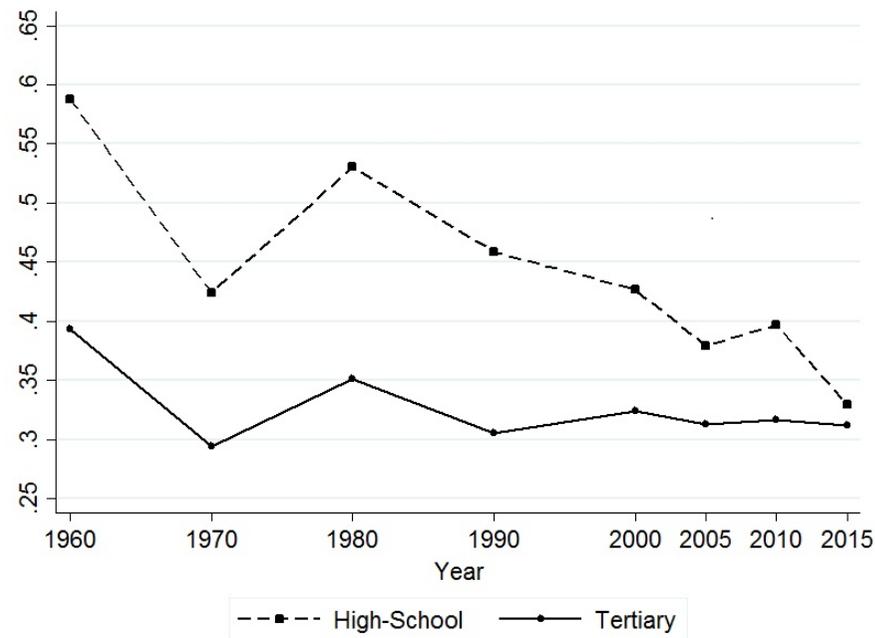

**a) young adult–mother**  **b) young adult–father**

**Figure 1. The dynamics of the Liu–Lu measure for two different classifications of the educational attainments (US 1960-2015).**

*Source*: authors' calculation using data from the international version of Integrated Public Use Microdata Series (IPUMS) from the Minnesota Population Center.

*Notes*: (i) For the dashed lines, the education level is considered to be "low" if it is lower than completed *high school* and it is classified to be "high" if at least a *high school* degree was obtained. (ii) For the solid lines, the education level is considered to be "low" if it is lower than *tertiary education* with a diploma and it is classified to be "high" if at least a *college degree* was obtained. (iii) Young adults are observed at the age between 30 and 40 years. (iv) We do not construct conventional confidence intervals, because there is hardly any sampling variation in our large and representative census data.

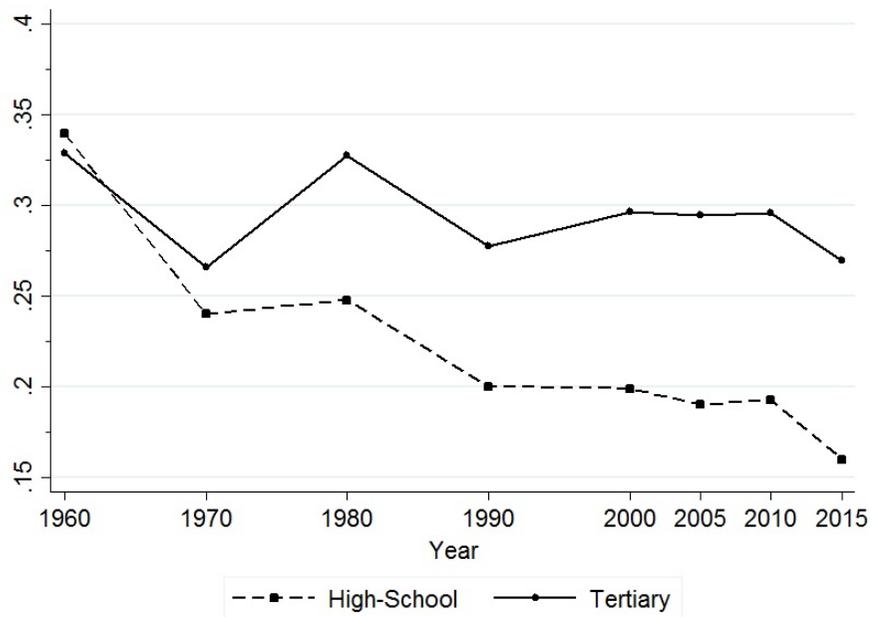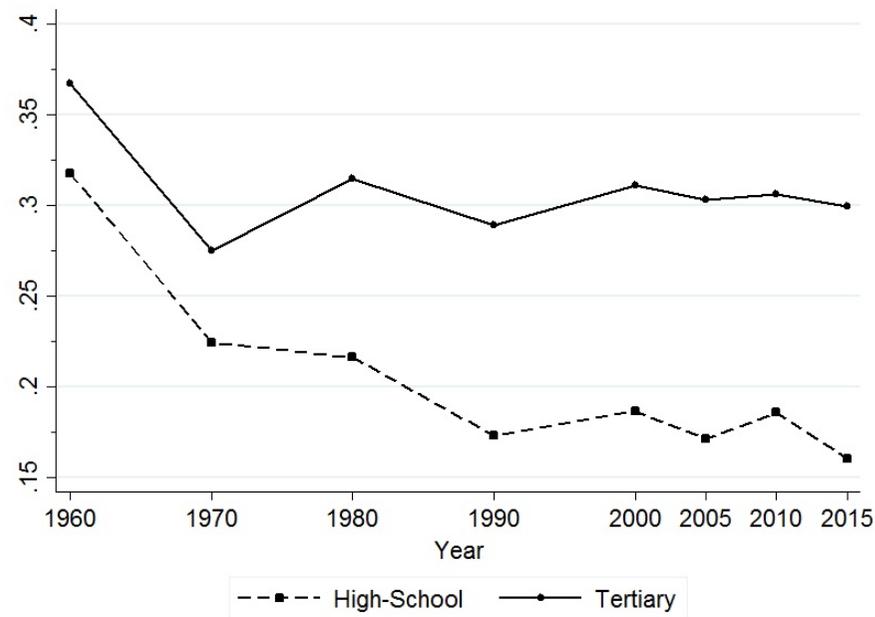

a) young adult–mother  b) young adult–father

**Figure 2. The dynamics of the intergenerational persistence coefficient for two different classifications of the educational attainments (US 1960-2015).**

*Notes*: same as under Fig.1. The IGPC is the $\beta$ coefficient in either of the regression equations of

$edattain_i = \alpha + \beta\, edattain_{mom,i} + \epsilon_i$ and $edattain_i = \alpha + \beta\, edattain_{pop,i} + \epsilon_i$..

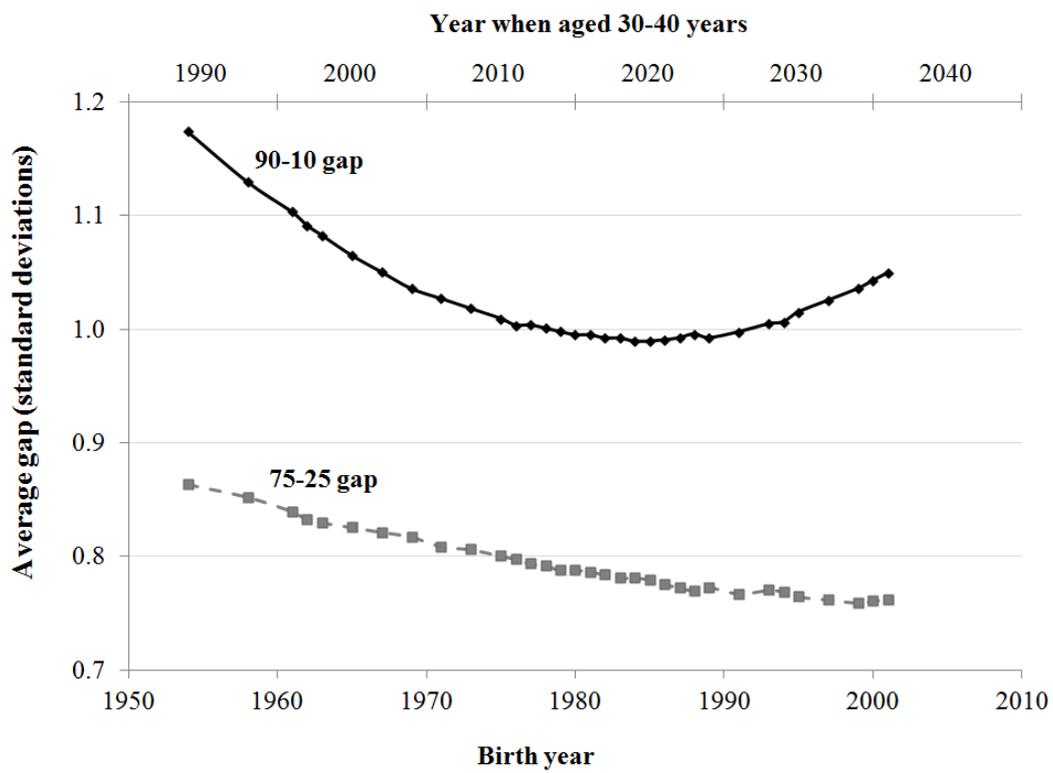

**Figure 3. Difference in test performance of American students by socioeconomic status.**

*Source*: Replication of Fig.1 in Hanushek et al. (2019).

*Notes*: The solid line represents average test scores of students whose family SES is at the 90th percentile of the income distribution minus that of students whose family SES is at the 10th percentile. The dashed line represents average test scores of students whose family SES is at the 75th percentile of the income distribution minus that of students whose family SES is at the 25th percentile. Students were born between 1954 and 2001. Family SES is measured via parents' education and student reports of home items. Test subjects are reading, math, and science. The students sit the tests when aged 14 to 17 years. Number of students is 2.7 million. The tests are the Long-Term Trend NAEP, Main NAEP, PISA, and TIMSS.

# Online appendix of the paper

# "Are high school degrees and university diplomas equally heritable in the US?

# A new measure of relative intergenerational mobility"

In our paper, we find that the Liu–Lu measure and the intergenerational persistence coefficient (IGPC) generate different rankings with respect to the inequality of opportunities associated to parents' education. Specifically, if the studied inequality of opportunity is quantified by the Liu–Lu measure, then it is higher when the examined opportunity relates to completing *high school* compared to the case when it is related to achieving a *tertiary education diploma*.

While in our paper, we use data exclusively for the United States, in this online appendix we expand the analysis to other countries as well. We show that the sensitivity of the ranking to the applied measure is not specific to the US. By examining the census data of Austria, France, Greece, Hungary, Portugal, Romania, and Spain, we find at least one year over the period between 1968 and 2011 for each of these countries, when the ranking depends on how mobility is captured. Moreover, we show that this sensitivity does not disappear for the US if the IGPC is replaced by another conventional measure, the correlation coefficient.

Fig 1. The dynamics of the Liu–Lu measure and the IGPC for two classifications of the educational attainments between 1960 and 2015 in the **USA**

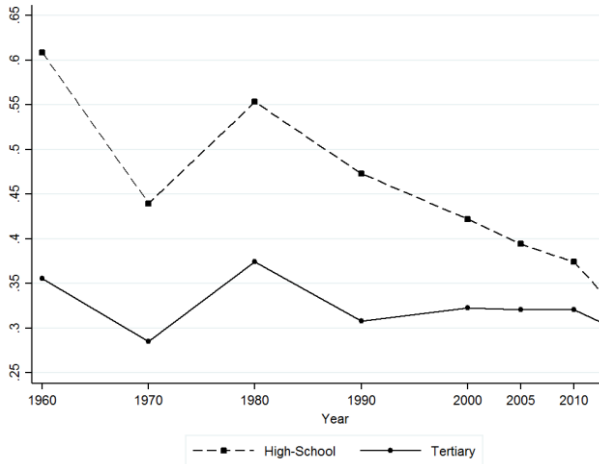

(a) Liu–Lu measure
young adult–mother

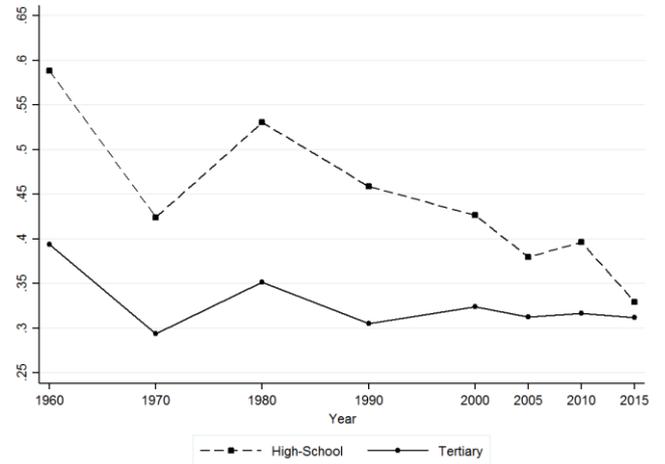

(b) Liu–Lu measure
young adult–father

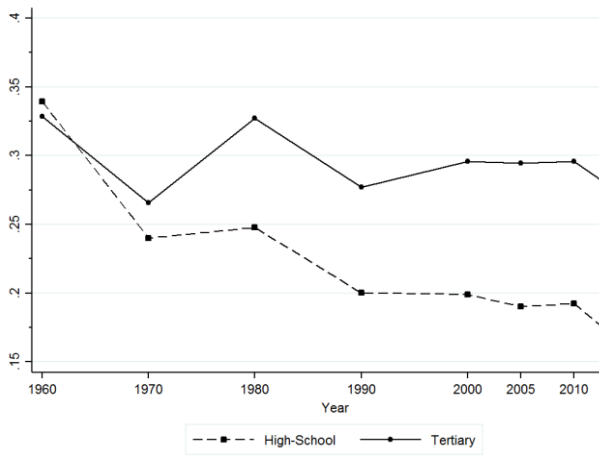

(c) Intergenerational persistence coefficient
young adult–mother

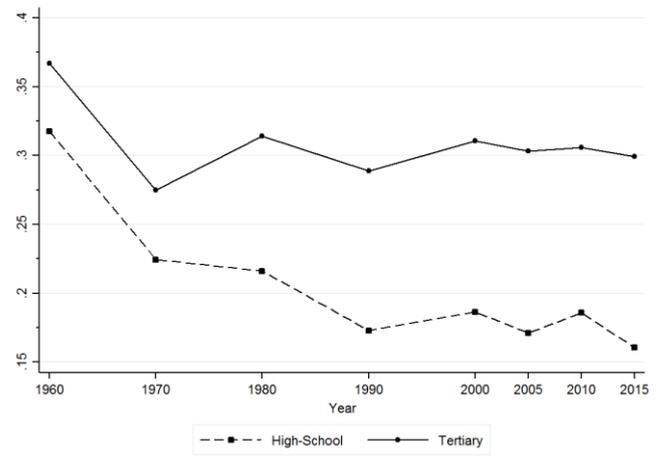

(d) Intergenerational persistence coefficient
young adult–father

*Source*: replication of Figures 1, and 2 in our paper by using US census data from IPUMS. We use the variable on the reported highest educational attainments of the respondents ($edattain$), their mother ($edattain_{mom}$) and their father ($edattain_{pop}$).

*Notes*: (i) For the dashed lines, the education level is considered to be "low" if it is lower than completed *high school* and it is classified to be "high" if at least a *high school* degree was obtained. (ii) For the solid lines, the education level is considered to be "low" if it is lower than *tertiary education* with a diploma and it is classified to be "high" if at least a *college degree* was obtained. (iii) Young adults are aged between 30 and 40 years when being interviewed.

**Main finding**: the ranking is sensitive to the choice of the measure in the years 1970, 1980, 1990, 2000, 2005, 2010, and 2015. If the studied inequality of opportunity is quantified by the Liu–Lu measure (Intergenerational persistence coefficient), then it is higher (lower) when the examined opportunity relates to completing high school compared to the case when it is related to achieving a tertiary education.

Fig 2. The dynamics of the Liu–Lu measure and the IGPC for two classifications of the educational attainments between 1971 and 2001 in **Austria**

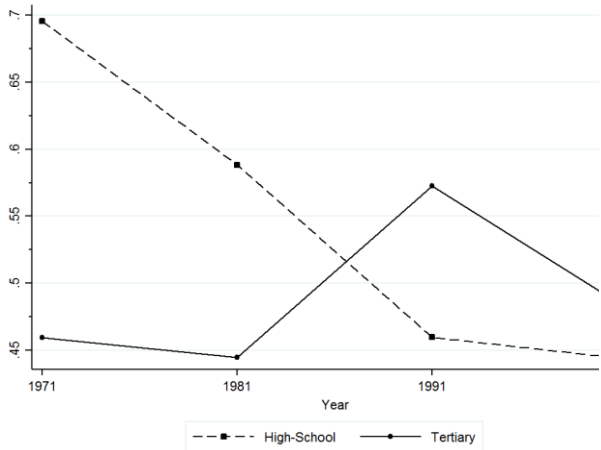

(a) Liu–Lu measure
young adult–mother

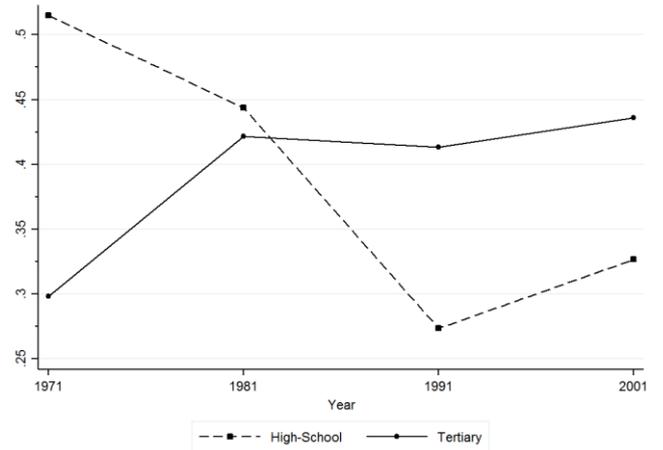

(b) Liu–Lu measure
young adult–father

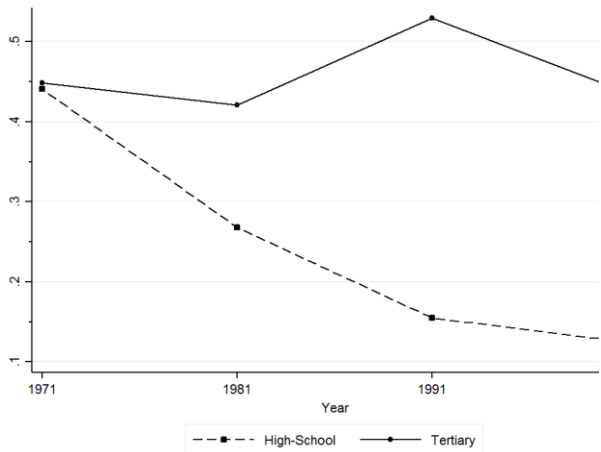

(c) Intergenerational persistence coefficient
young adult–mother

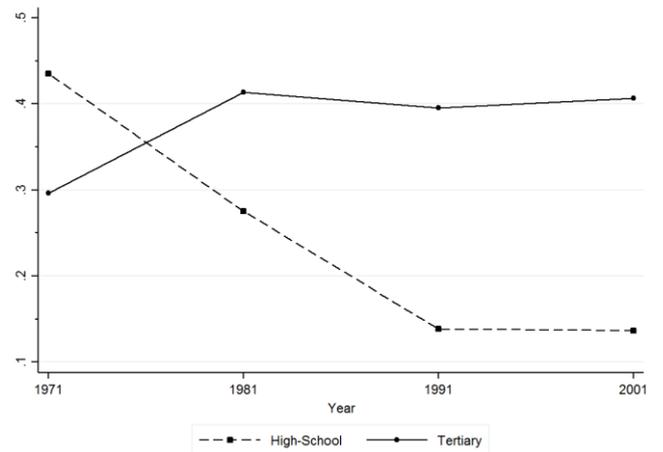

(d) Intergenerational persistence coefficient
young adult–father

**Main finding**: the ranking is reversed in the year 1981.

Fig 3. The dynamics of the Liu–Lu measure and the IGPC for two classifications of the educational attainments between 1968 and 2011 in **France**

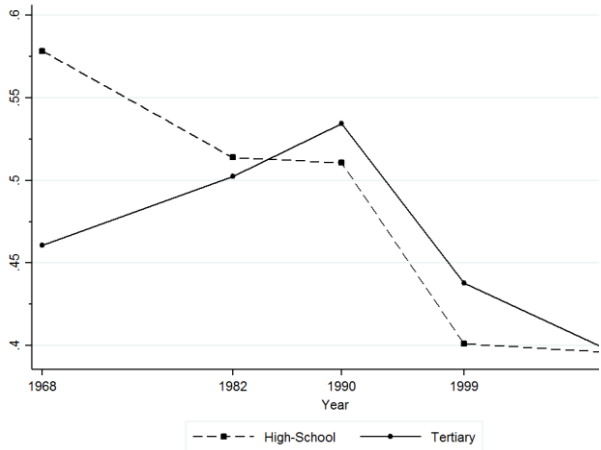
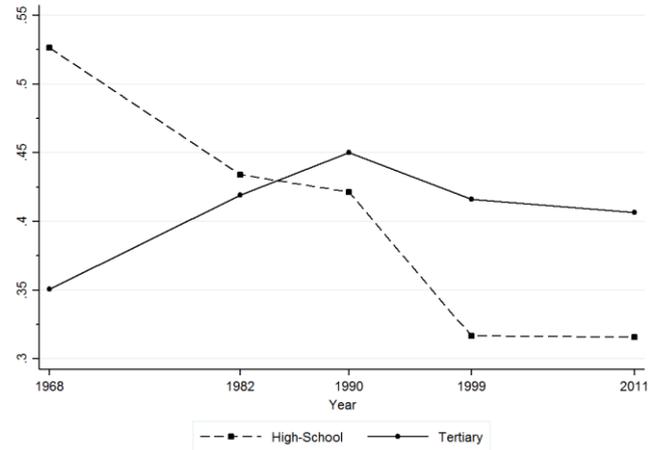

(a) Liu–Lu measure
young adult–mother

(b) Liu–Lu measure
young adult–father

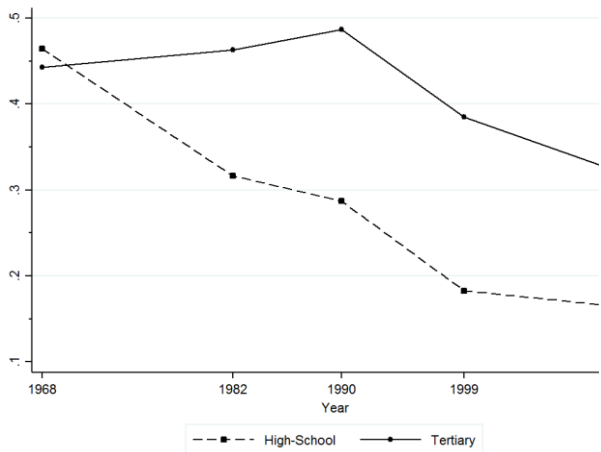
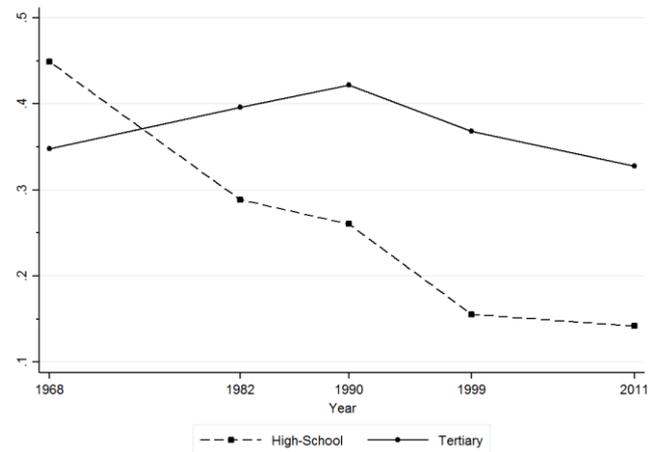

(c) Intergenerational persistence coefficient
young adult–mother

(d) Intergenerational persistence coefficient
young adult–father

**Main finding**: the ranking is reversed in the year 1982.

Fig 4. The dynamics of the Liu–Lu measure and the IGPC for two classifications of the educational attainments between 1971 and 2001 in **Greece**

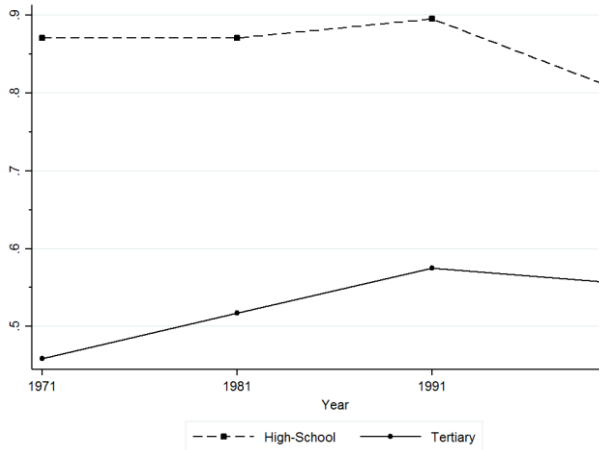
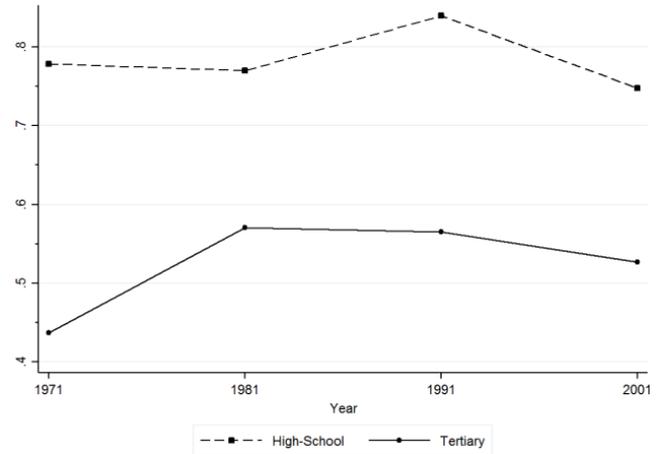

(a) Liu–Lu measure
young adult–mother

(b) Liu–Lu measure
young adult–father

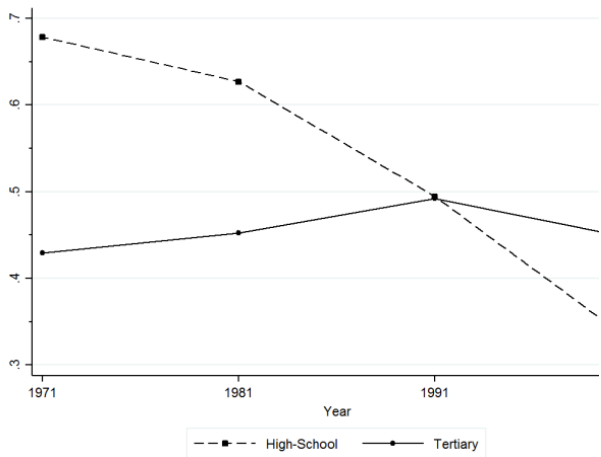
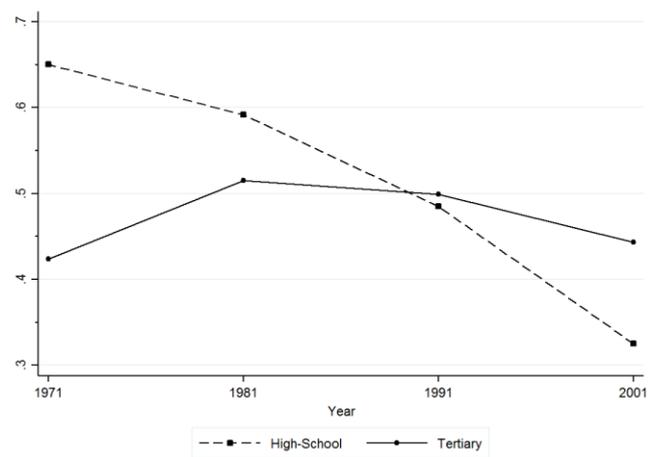

(c) Intergenerational persistence coefficient
young adult–mother

(d) Intergenerational persistence coefficient
young adult–father

**Main finding**: the ranking is reversed in the year 2001.

Fig 5. The dynamics of the Liu–Lu measure and the IGPC for two classifications of the educational attainments between 1970 and 2011 in **Hungary**

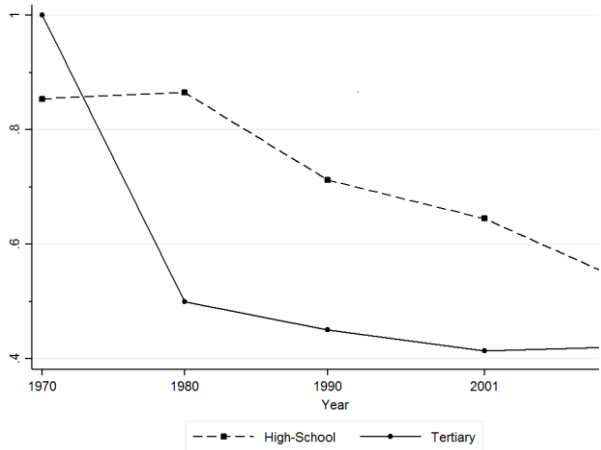

(a) Liu–Lu measure
young adult–mother

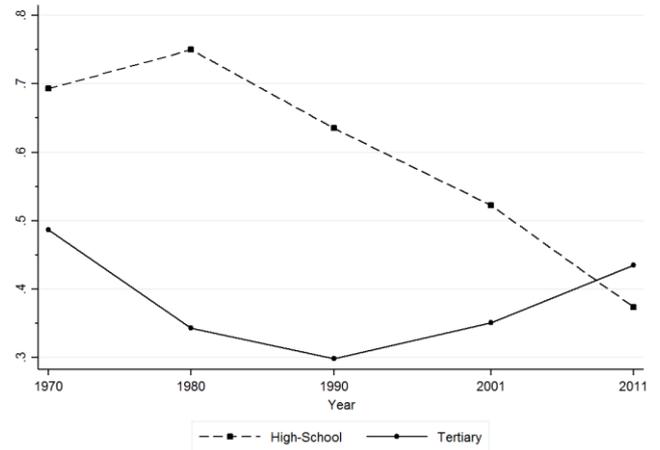

(b) Liu–Lu measure
young adult–father

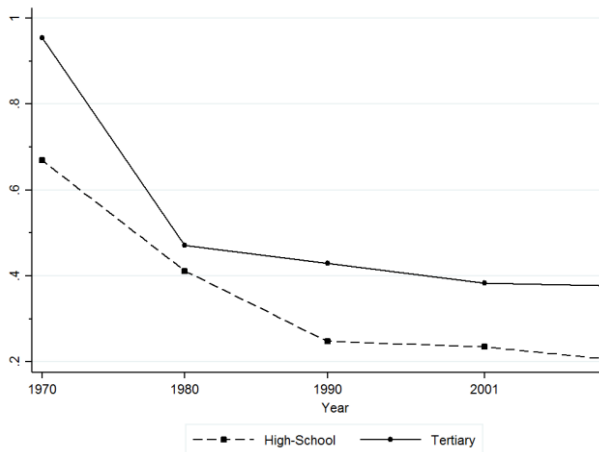

(c) Intergenerational persistence coefficient
young adult–mother

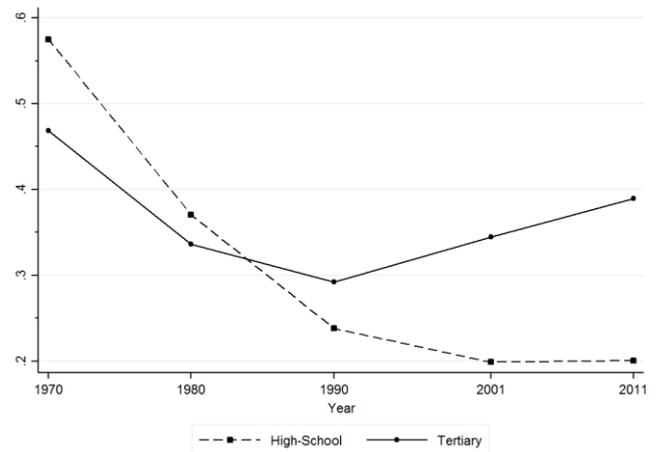

(d) Intergenerational persistence coefficient
young adult–father

**Main finding**: the ranking is reversed in the years 1990, and 2001.

Fig 6. The dynamics of the Liu–Lu measure and the IGPC for two classifications of the educational attainments between 1981 and 2011 in **Portugal**

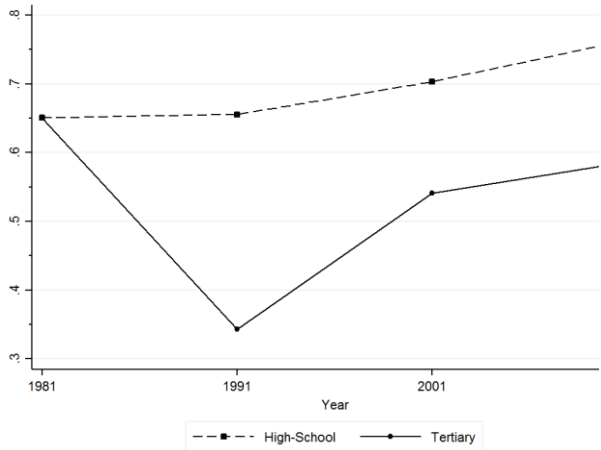

(a) Liu–Lu measure
young adult–mother

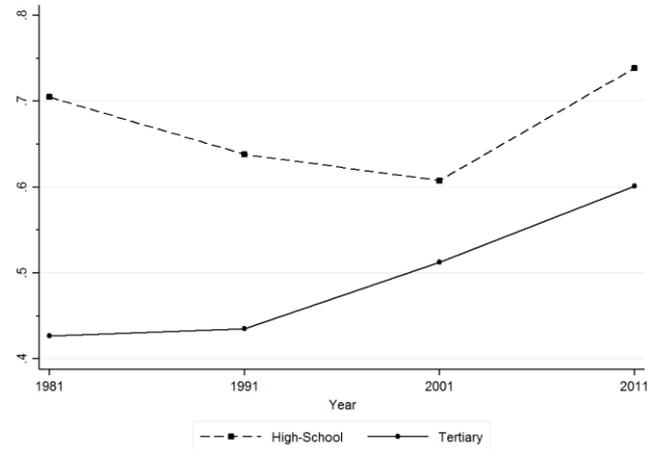

(b) Liu–Lu measure
young adult–father

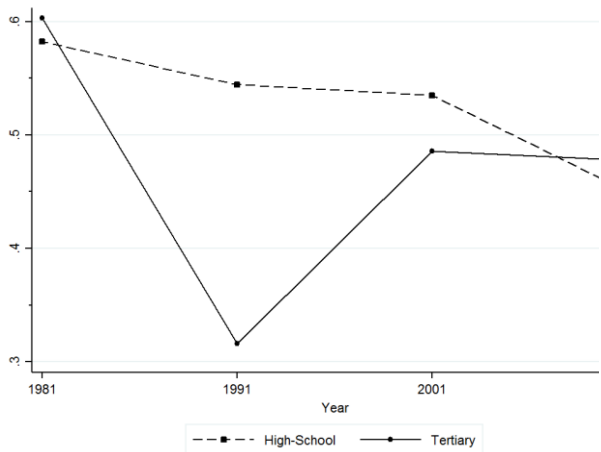

(c) Intergenerational persistence coefficient
young adult–mother

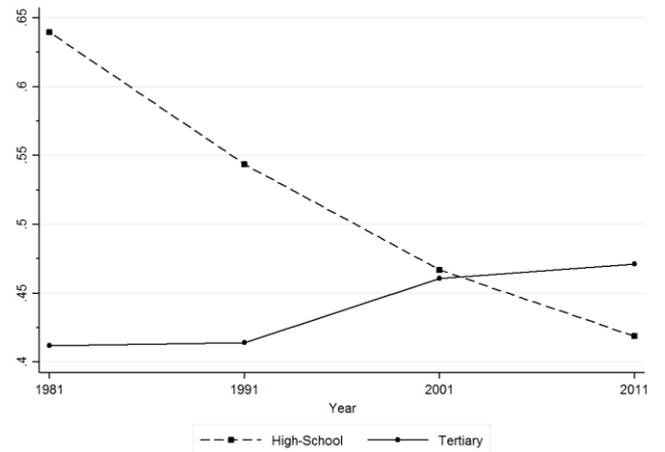

(d) Intergenerational persistence coefficient
young adult–father

**Main finding**: the ranking is reversed in the year 2011.

Fig 7. The dynamics of the Liu–Lu measure and the IGPC for two classifications of the educational attainments between 1977 and 2011 in **Romania**

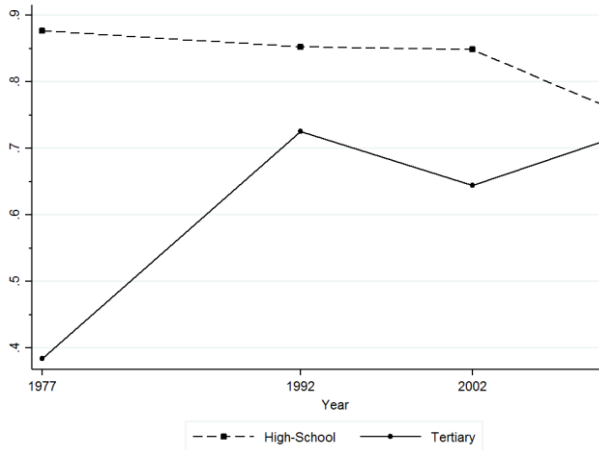

(a) Liu–Lu measure
young adult–mother

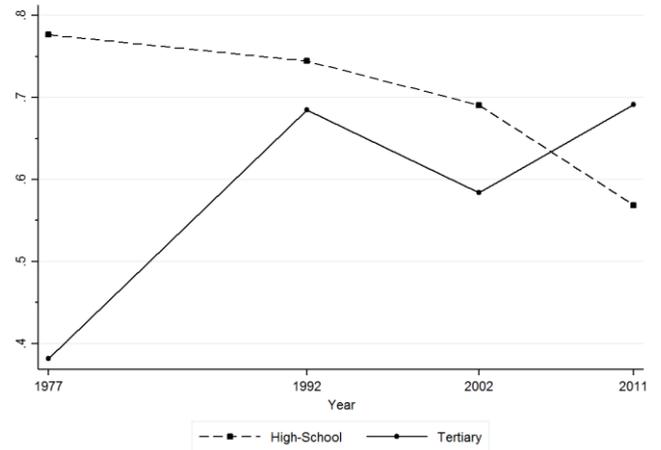

(b) Liu–Lu measure
young adult–father

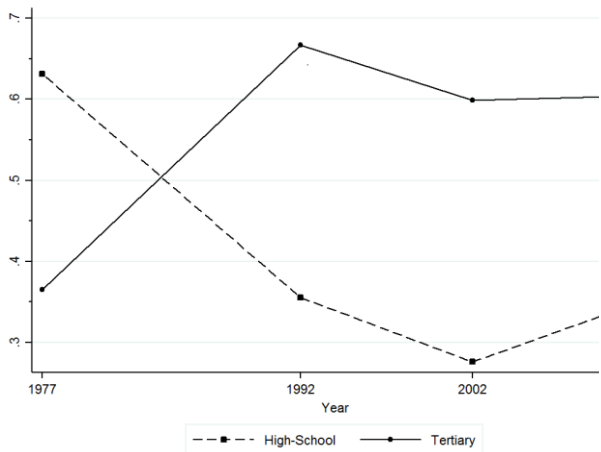

(c) Intergenerational persistence coefficient
young adult–mother

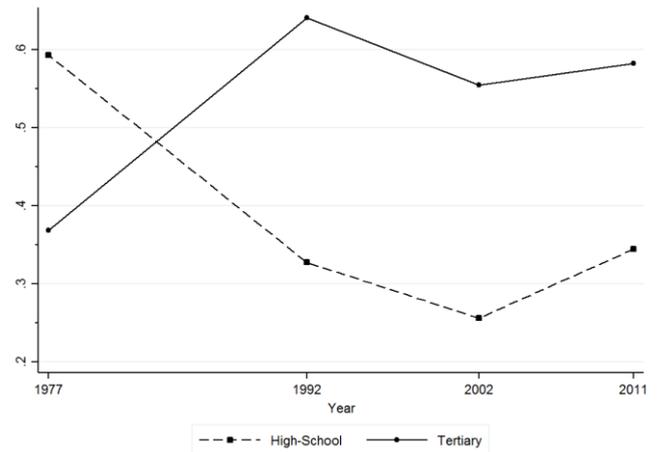

(d) Intergenerational persistence coefficient
young adult–father

**Main finding**: the ranking is reversed in the years 1992, and 2002.

Fig 8. The dynamics of the Liu–Lu measure and the IGPC for two classifications of the educational attainments between 1991 and 2011 in **Spain**

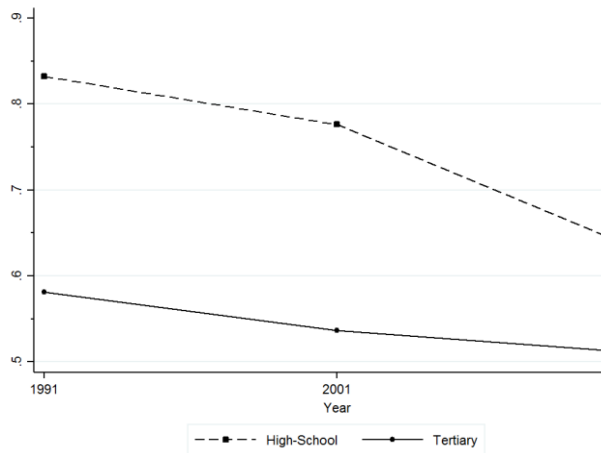
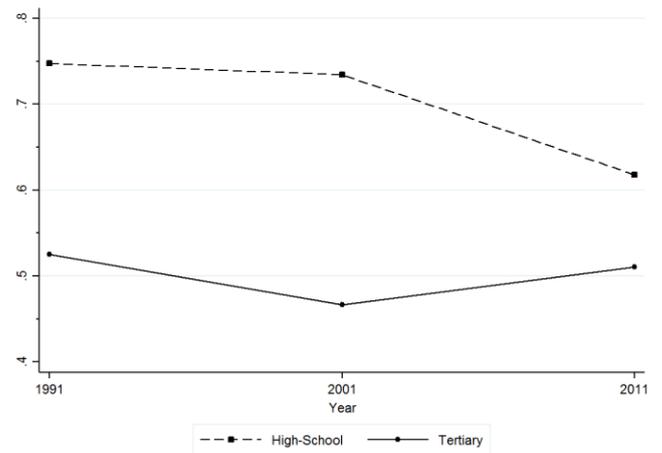

(a) Liu–Lu measure
young adult–mother

(b) Liu–Lu measure
young adult–father

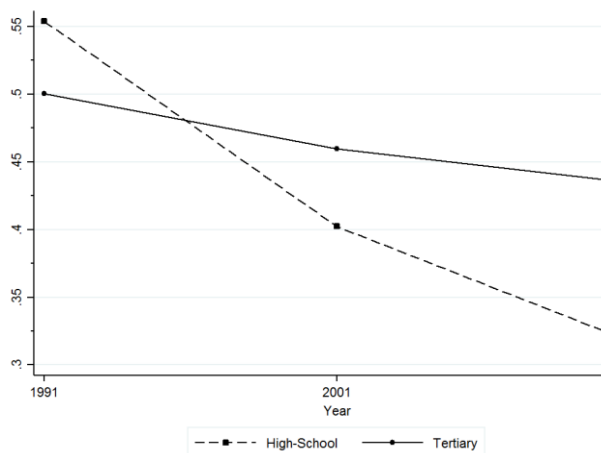
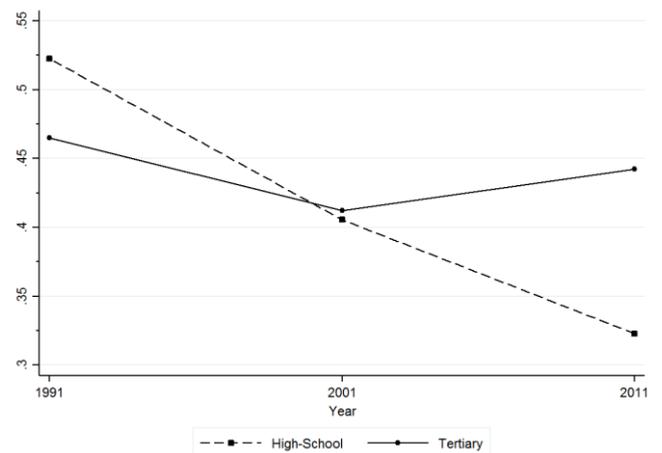

(c) Intergenerational persistence coefficient
young adult–mother

(d) Intergenerational persistence coefficient
young adult–father

**Main finding**: the ranking is reversed in the years 2001, and 2011.

Fig 9. The dynamics of the Liu–Lu measure and the **correlation coefficient** for two classifications of the educational attainments between 1960 and 2015 in the **USA**

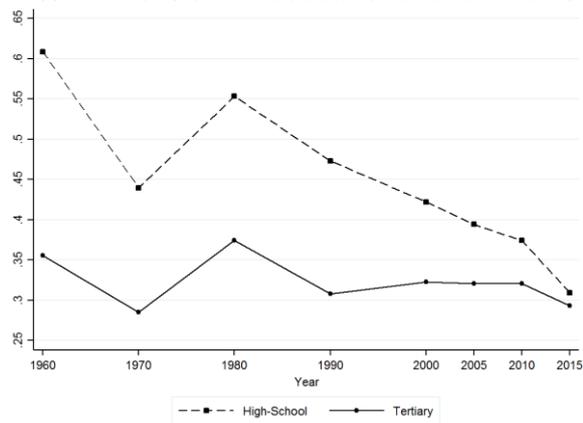

(a) Liu–Lu measure
young adult–mother

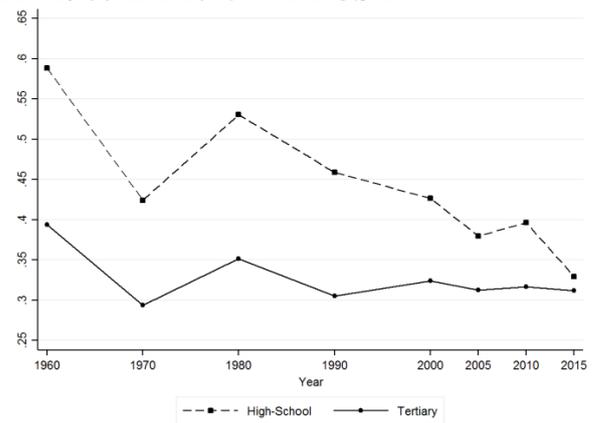

(b) Liu–Lu measure
young adult–father

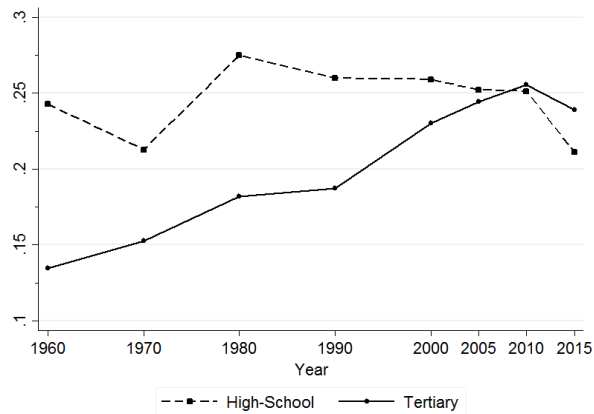

(c) Correlation coefficient
young adult–mother

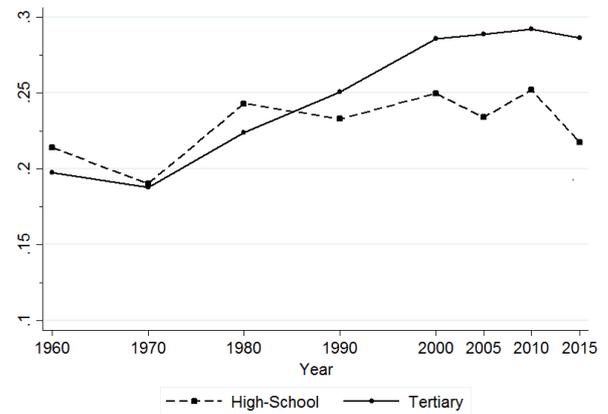

(d) Correlation coefficient
young adult–father

**Main finding**: the ranking is reversed in the years 2010, and 2015.